\documentclass[12pt,english,floatfix,superscriptaddress,aps,prd,preprint,showkeys,nofootinbib]{revtex4}
\usepackage{amsmath}
\usepackage{amssymb}
\usepackage{amsbsy}
\usepackage{amsfonts}
\usepackage{amsopn}
\usepackage{amstext}
\usepackage{graphicx}
\usepackage{amssymb}
\usepackage{amsfonts}
\usepackage{amsmath}
\usepackage{graphicx}
\usepackage[english]{babel}
\usepackage{color}
\usepackage{slashed}
\usepackage{esint}
\usepackage[dvips]{epsfig}
\usepackage[dvips]{graphicx}
\usepackage{float}
\usepackage{units}
\usepackage{textcomp}
\newcommand{\beq}{\begin{equation}}
\newcommand{\eeq}{\end{equation}}
\newcommand{\bea}{\begin{eqnarray}}
\newcommand{\eea}{\end{eqnarray}}

\begin{document}

\title{Casimir Wormholes in $2+1$ Dimensions with Applications to the Graphene}

\author{G. Alencar \footnote{geova@fisica.ufc.br}}

\affiliation{${}^{*}$\!Universidade Federal do Cear\'a, Fortaleza-CE, Brazil.}
\affiliation{${}^{}$\!International Institute of Physics - Federal University of Rio Grande do Norte, Campus Universit\'ario, Lagoa Nova, Natal, RN 59078-970, Brazil.}

\author{V. B. Bezerra\footnote{E-mail:valdir@fisica.ufpb.br}}

\affiliation{${}^{\dag}$\!Universidade Federal da Para\'iba, Departamento de F\'isica, Jo\~ao Pessoa, PB,Brazil.}

\author{C. R. Muniz\footnote{E-mail:celio.muniz@uece.br}}

\affiliation{${}^{\ddag}$\!Universidade Estadual do Cear\'a, Faculdade de Educa\c c\~ao, Ci\^encias e Letras de Iguatu, Iguatu-CE, Brazil.}



\begin{abstract}

\vspace{0.75cm}
In this paper we show that wormholes in (2+1) dimensions (3-D) cannot be sourced solely by both Casimir energy and tension, differently from what happens in a 4-D scenario, in which case it has been shown recently, by the direct computation of the exact shape and redshift functions of a wormhole solution, that this is possible. We show that in a 3-D spacetime the same is not true since the arising of at least an event horizon is inevitable. We do the analysis for massive and massless fermions, as well as for scalar fields, considering quasi-periodic boundary conditions and find that a possibility to circumvent such a restriction is to introduce, besides the 3-D Casimir energy density and tension, a cosmological constant, embedding the surface in a 4-D manifold and applying a perpendicular weak magnetic field. This causes an additional tension on it, which contributes to the formation of the wormhole. Finally, we discuss the possibility of producing the condensed matter analogous of this wormhole in a graphene sheet and analyze the electronic transport through it.
\end{abstract}

\renewcommand{\thesection}      {\Roman{section}}
\maketitle
\newpage
\section{INTRODUCTION}
\indent Wormholes originally are solutions to the field equations of General Relativity that show unexpected connections between two quite separated regions of the spacetime \cite{EinsteineRose,Wheeler,Ellis,Visser}, occurring even in D-dimensional spacetimes and with several topologies (\cite{Lemos}, and references therein). They do not satisfy the energy conditions of the General Relativity, being necessary some type of exotic matter as source, with some exceptions \cite{wormholecasimir1, wormholecasimir2, wormholecasimir3, wormholecasimir4}. Thus, the Casimir effect, that generally involves negative energies of free quantum fields subject to certain boundary conditions, has been increasingly examined in the context of wormholes \cite{Maldacena1}. Moreover, the study of the relationship between the Casimir effect and traversable wormholes can lead to the arising of novel insights with respect to the issue if gravity in fact influences the vacuum energy (and, vice-versa, if this latter gravitates), at least in a weak field regime. This  topic is actually object of discussion \cite{Geova,Sorge2} as well as of projects for observational investigations, as in the Archimedes experiment \cite{Calloni}.

Recent works considering the Casimir effect in space-times around of wormholes have been published \cite{Sorge,Alana,Bezerra}, as well as others which analyze how traversable wormholes can be produced and sustained by means of both the Casimir energy and tension, in the context of General Relativity and extended theories of gravitation, in semiclassical approaches  \cite{Remo,Jusufi,Sunil}. In these works it has been demonstrated that in a 4-D spacetime such quantities are feasible sources to a Morris-Thorne wormhole from the direct calculation of the redshift and shape functions associated to this object. In the present paper, we will investigate 3-D traversable wormholes and show that this construction is not possible, since at least an event horizon appears when one considers only the Casimir quantities as gravity source.

We will do this analysis by considering massive and massless fermions, as well as scalar fields, adopting quasi-periodic boundary conditions. We will overcome the aforementioned restriction concerning 3-D Casimir wormholes by introducing a cosmological constant (which corresponds to a preexisting tension on the surface under investigation), embedding the 3-D surface in a 4-D manifold and applying a weak uniform magnetic field perpendicularly to the surface. We then will apply the model for a graphene sheet, since a fermion on it exhibits a simulacrum of relativistic behavior \cite{Castro}, obtaining thus an asymptotically conical wormhole by taking into account anti-periodic boundary conditions for the fermion coupled to the external magnetic field. Furthermore, we will study the conditions for the electronic transport to occur throughout the wormhole, comparing with the carriers motion through a flat sheet. In this sense, our propose differs from the ones discussed in \cite{Gonzales,Furtado}, which did not analyze the role played by the Casimir energy and tension in the graphene wormhole, since it seems to exist a relation of dependence between this latter and those quantities, as already discussed.

The manuscript is organized as follows. In section II we show that the usual Casimir energy of a massless fields, solely, can not be a source of a wormhole. We add a cosmological constant and other general sources as a solution. In section III we study if the addition of mass or quasi-periodic boundary conditions to the Casimir energy can generate the source pointed out in section II. In section IV we consider a graphene sheet and show that a perpendicular magnetic field can solve the problem. We also discuss some phenomenological consequences. Finally, in section V we present our concluding remarks.

\section{Traversable Casimir wormholes in $(2+1)$ dimensions}

In this section we analyze if it is possible, as in the 4-D case, to sustain a traversable wormhole in a 3-D spacetime from the Casimir quantities, namely, energy density and tension. Initially, we take the general metric of a traversable circularly symmetric 3-D wormhole, according to \cite{Mann}
\begin{equation}
\displaystyle ds^2 = - e^{2\Phi(r)}c^2dt^2 + \frac{dr^2}{1-b(r)/r} + r^2 d\phi^2,   \label{eq1}
\end{equation}
where $\Phi(r)$ and $b(r)$ are the redshift and shape functions, respectively. Einstein's equations in an orthonormal basis are, therefore
\begin{eqnarray}\label{EinEq}
G_{tt}&=&\frac{b'r-b}{2 r^3}=\kappa \rho(r),\nonumber \\
G_{rr}&=&-\frac{r-b}{r^2}\Phi'= \kappa \tau(r)\nonumber \\
G_{\phi\phi}&=&\frac{(r-b)}{r}\left[\Phi''-\frac{(b'r-b)\Phi'}{2r(r-b)}+(\Phi')^2\right]=\kappa p(r),
\end{eqnarray}
where (') means the derivative with respect to $r$; $\rho(r)$ is the surface energy density, $\tau(r)$ and $p(r)$ the radial and transverse tensions, respectively. The Einstein constant is $\kappa = 8\pi G c^{-4}$, where $G$ is the gravitational constant and $c$ is the light velocity. The first thing we should point about the above equations is that they are quite different from the 4-D case.

According to the first of Eqs. (\ref{EinEq}), the flare out condition valid for the wormhole, $b'r-b<0$, just is obeyed if $\rho(r)<0$. The Casimir apparatus is a typical example of a system with negative energy, and we will use this fact in order to build our wormhole, by following Ref. \cite{Remo}. The Casimir energy density of a massless field in a 3-D spacetime is usually given by the expression
\begin{equation}\label{DensEn}
\rho_C(r)=-\frac{\lambda}{r^3},
\end{equation}
where $\lambda$ will depend on the specific case considered. A first result here is that the Casimir energy density obtained from $\lambda>0$, which is positive, does not generate wormholes, since the flare out condition is not satisfied.

The Casimir radial tension is given by
\begin{equation}\label{RadTen}
\tau_C(r)=-2\frac{\lambda}{ r^3},
\end{equation}
so that the Equation of State(EoS) is $\tau_C=2\rho_C$. This non-zero quantity indicates that the redshift function cannot be a constant (as $\Phi=0$, which would give a zero tidal wormhole), according to Eqs. (\ref{EinEq}). Now we will substitute  Eq.(\ref{DensEn}) into the first of the Eqs. (\ref{EinEq}) in order to determine $b(r)$. Thus, we find that
$$
b=\frac{r_{0}-2\kappa \lambda}{r_{0}}r+2\kappa \lambda.
$$
The constant of integration was fixed such that $b(r_0)=r_0$, where $r_0$ is the throat of the wormhole. Now by using this and  Eq.(\ref{RadTen}) into the second of Eqs. (\ref{EinEq}), we determine $\Phi(r)$, which is be given by
$$
\Phi=\Phi_{0}(r_{0})+\ln(r-r_0)-\ln(r).
$$

Choosing the constant $\Phi_0$ equal to zero, we get the simple solution
$$
\Phi=\ln\left(1-\frac{r_{0}}{r}\right).
$$
Finally, by using the above results we arrive at the metric
\begin{equation}\label{WormMet}
ds^2 = - \left(1-\frac{r_0}{r}\right)^2c^2dt^2 + \frac{ r_0}{2\kappa  \lambda (1-\frac{r_0}{r})}dr^2 + r^2 d\phi^2,
\end{equation}
Unfortunately, this solution does not represent in fact a wormhole, since there exists a horizon at $r=r_0$. This is very different from the 4-D case, where the introduction of the tidal effect was enough to provide a consistent Casimir wormhole \cite{Remo}. Thus, at least with the usual 3-D Casimir energy and tension, it is not possible to generate a wormhole in such a spacetime. In what follows we will analyze some possibilities to solve this.

In order to circumvent the pointed problem, we add modifications to both the Casimir energy and radial tension, given by
\begin{equation}\label{constant}
\rho_C(r)=\lambda_0+\frac{\lambda_{1}}{r}+\frac{\lambda_{2}}{r^2}-\frac{\lambda}{r^3} , \tau_C(r)=-\lambda_0-2\frac{\lambda}{r^3}.
\end{equation}
The origin of $\lambda_0,\lambda_1,\lambda_1,$ will be analyzed latter. We should point out that the above quantities do not satisfy $\tau_C(r)=2\rho_C(r)$ anymore.   We will also introduce a cosmological constant $\Lambda$, which can be seen as a tension on the surface. Now, we seek for a metric in the form \cite{Mann2}
\begin{equation}
ds^2=-e^{2\Phi}dt^2+\frac{dr^2}{\Lambda r^2-M(r)}+r^2d\phi^2,
\end{equation}
with
\begin{eqnarray}
G_{tt}-\Lambda g_{tt}&=&\frac{M'}{2 r}=\kappa \rho_C(r) \label{EinEq2}\\
G_{rr}-\Lambda g_{rr}&=&\left[\Lambda-\frac{\Phi'}{r}(\Lambda r^2-M)\right]=\kappa\tau_C(r) \label{EinEq3} \\
G_{\phi\phi}-\Lambda g_{\phi\phi}&=&(\Lambda r^2-M)\left[\Phi''+\frac{(2\Lambda r-M')\Phi'}{2(\Lambda r^2-M)}+(\Phi')^2\right]-\Lambda=\kappa p_C(r).\label{EinEq4}
\end{eqnarray}
After substituting the new Casimir energy density, Eq. (\ref{constant}), into Eq. (\ref{EinEq2}) we find
\begin{equation}
M(r)=M_0+\kappa\lambda_{0}r^{2}+2\kappa\lambda_{1}r+2\kappa\lambda_{2}\ln r+2\kappa\frac{\lambda}{r}.
\end{equation}
Considering that the space must be asymptotically flat when $r\to\infty$, then we will impose
\begin{equation}\label{flatcondition}
\Lambda=\kappa\lambda_0.
\end{equation}
Hence, we get
\begin{equation}
g_{rr}^{-1}=-2\kappa\lambda_{1}(r-r_{0})-2\kappa\lambda_{2}\ln\frac{r}{r_{0}}-2\kappa\lambda(\frac{1}{r}-\frac{1}{r_{0}}),
\end{equation}
which is equals to the one found in Eq. (\ref{WormMet}) when $\lambda_{1}=0=\lambda_{2}$.

In what follows, we will determine the redshift function, $\Phi(r)$, by solving Eq.(\ref{EinEq2}) with the tension corrected and the fixed value for $\Lambda$. In order to find analytical solutions we consider the simplified case $\lambda_{2}=0$. With this we find two simple solution, namely,
\begin{equation}
\Phi_1=\frac{\Lambda r_{0}^{2}}{\kappa\lambda}r+\frac{\Lambda r_{0}}{2\kappa\lambda}r^{2}+\left(\frac{\Lambda r_{0}^{3}}{\kappa\lambda}+1\right)\ln(r-r_{0})-\ln \frac{r}{r_0},
\end{equation}
for $\lambda_{1}=0$ and

\begin{equation}
\Phi_2=-\frac{\Lambda}{\kappa\lambda_{1}}r-\frac{\lambda}{(\lambda_{1}r_{0}^{2}-\lambda)}\left(\frac{r_{0}^{3}\Lambda}{\kappa\lambda}+1\right)\ln(r-r_{0})-c_1\ln(r-\frac{\lambda}{r_{0}\lambda_{1}})-\ln \frac{r}{r_0},
\end{equation}
for $\lambda_{1}\neq0$, where
$$
c_1=\frac{\lambda^{2}\Lambda+\kappa\lambda_{1}^{3}r_{0}^{3}}{r_{0}\lambda_{1}^{2}\kappa(\lambda-\lambda_{1}r_{0}^{2})}.
$$
The integration constants are fixed in order to leave the logarithm argument without dimension. Now we analyze the conditions to avoid an event horizon. For both solutions we see that we must impose
\begin{equation} \label{horizon}
\frac{\Lambda }{\kappa\lambda}=-\frac{1}{r_{0}^{3}} .
\end{equation}
For $\Phi_2$ we must impose two further conditions
\begin{eqnarray} \label{horizon2}
\frac{\lambda}{\lambda_{1}}&<&0,\nonumber\\
\frac{\Lambda}{\lambda_{1}}&>&0.
\end{eqnarray}
The first is in order to avoid the event horizon, and the second that the metric does not diverge at infinity. We finally get
the final wormhole metrics
\begin{equation}
ds_1^2=-\exp{\left[-\frac{2r}{r_0}-\frac{r^2}{r_0^2}\right]\left(\frac{r_0}{r}\right)^2}dt^2+\frac{r_0}{2\kappa \lambda(1-\frac{r_0}{r})}dr^2 + r^2 d\phi^2
\end{equation}
and
\begin{equation}
ds_2^2=-e^{-\frac{2\Lambda}{\kappa\lambda_{1}}r}\frac{1}{(r-\frac{\lambda}{r_{0}\lambda_{1}})^{2c_{1}}}(\frac{r_{0}}{r})^{2}dt^2-\frac{r}{2\kappa\lambda_{1}}\frac{1}{(r-r_{0})(r-\frac{\lambda}{r_{0}\lambda_{1}})}dr^2 + r^2 d\phi^2
\end{equation}
As a final conclusion we note that Eq. (\ref{horizon}), together with Eq. (\ref{flatcondition}), give us the relation
\begin{equation}
\lambda_0=-\frac{\lambda}{r_0^{3}}.
\end{equation}
Since $r_0>0$, we conclude that  $\lambda_0>0$. Beyond this, with (\ref{horizon2}) we also find that $\lambda_1<0$. Therefore, the signal of additional sources are completely fixed in order to get a wormhole solution. In the next sections we will consider the possible sources for $\lambda_0,\lambda_1$.

\section{Casimir energy with quasi-periodic boundary conditions}
In this section we look for some possibilities in order to get the extra terms in the energy density. We will consider scalar and fermion fields in $(d+1)$ spacetime dimensions. The standard procedure to obtain the Casimir energy is to consider periodic or anti-periodic boundary conditions, given by
\begin{equation}
\phi(t,\vec{x}+\vec{L})=\pm\phi(t,\vec{x}).
\end{equation}
However, in order to consider more general materials, metals or semimetals, for example, the authors of Ref. \cite{Feng:2013zza}  considered the general case
\begin{equation}
\phi(t,\vec{x}+\vec{L})=e^{i2\pi\theta}\phi(t,\vec{x}),
\end{equation}
where $0\leq\theta\leq 1$. In this way it is possible to consider, metallic($\theta=0$) or semimetallic ($\theta=\pm2\pi/3$) nanotubes. However the fermionic case considered in Ref. \cite{Suzuki} has not taken into account quasi-periodic conditions. Following Refs. \cite{Suzuki,Feng:2013zza}, we will obtain the Casimir energy for fermions and bosons with quasi-periodic boundary conditions.

\subsection{The Massless Case}
We first consider the massless case. The general boundary condition is given by
$$
\psi(t,\vec{x}+\vec{L})=e^{i2\pi\theta}\psi(t,\vec{x}),
$$
where $\psi$ is a general wave function. With the above condition, the spectrum is given by
$$
\omega_{n}^{2}=k_{T}^{2}+\left[\frac{2\pi(n+\theta)}{a}\right]^{2},
$$
and thus we can determine the density of energy, using the following relation
$$
\rho=p\frac{(-1)^{q}}{2a}\int_{-\infty}^{\infty}\frac{d^{d-1}k}{(2\pi)^{d-1}}\sum_{-\infty}^{\infty}\omega_{n}
$$
where $p$ accounts for the number of degrees of freedom of the field and $q=0,1$ for bosons and fermions respectively. Now, by using the result
$$
\int\frac{d^{d}k}{(2\pi)^{d}}(k^{2}+\Delta)^{-l}=\frac{1}{(4\pi)^{d/2}}\frac{\Gamma(l-\frac{d}{2})}{\Gamma(l)}\Delta^{\frac{d}{2}-l}.
$$
we get that, by taking $l=s/2$
$$
\rho=p(-1)^{q}\frac{\pi^{\frac{d+1}{2}}}{a^{d+1}}\frac{\Gamma(\frac{s+1-d}{2})}{\Gamma(\frac{s}{2})}\sum_{-\infty}^{\infty}(n+\theta)^{d-1-s}.
$$
Here we will follow a  path more direct than that used in ref. \cite{Feng:2013zza}.  In order to regularize the above expression we must note that the Epstein zeta function is given by
\begin{equation}\label{Epstein}
E(A,c,q,l)=\sum_{n}\left[\frac{A}{2}(n+c)^{2}+q\right]^{-l},
\end{equation}
and our expression becomes
$$
\rho=p(-1)^{q}\frac{\pi^{\frac{d+1}{2}}}{a^{d+1}}\frac{\Gamma(\frac{s+1-d}{2})}{\Gamma(\frac{s}{2})}E(A=2,c,q=0,l=\frac{s+1-d}{2}).
$$
However, Eq.(\ref{Epstein}) is valid only for $l>1/2$. In our case we need that $l=(s+1-d)/2<1/2$ and one could say that the above expression is useless for us. It is a known fact that Eq.(\ref{Epstein}) can be analytically continued into a meromorphic function in the whole complex plane\cite{Elizalde}. Therefore, after performing a Poisson resumation, we find
\begin{equation}\label{elizalde}
E(2,c,q,l)=\sqrt{\pi}q^{\frac{1}{2}-l}\frac{\Gamma(l-\frac{1}{2})}{\Gamma(l)}+\frac{2^{2}\pi^{l}q^{-\frac{l}{2}+\frac{1}{4}}}{\Gamma(l)}\sum_{n=1}^{\infty}\cos(2\pi nc)n^{(l-\frac{1}{2})}K_{\frac{1}{2}-l}\left(2\pi n\sqrt{q}\right).
\end{equation}
By using the above expression with
$$
c=\theta,q=0,l=(s+1-d)/2
$$
and by performing the limit $s\to-1$ we arrive at the general Casimir energy
\begin{equation}\label{casimirgeneral}
\rho=-p(-1)^{q}\frac{1}{a^{d+1}}\frac{\Gamma(\frac{d+1}{2})}{\pi^{\frac{d+1}{2}}}\sum_{n=1}^{\infty}\frac{\cos(2\pi n\theta)}{n^{d+1}}.
\end{equation}
Again, for $(p,q)=1,0$, we reobtain the same result found in Ref. \cite{Feng:2013zza} for the scalar field. However now we can consider other spins and arbitrary boundary conditions. For $d=2$ we get
\begin{equation}\label{CasimirScalar3D}
\rho=-p(-1)^{q}\frac{1}{2\pi a^{3}}\sum_{n=1}^{\infty}\frac{\cos(2\pi n\theta)}{n^{3}}.
\end{equation}
For $\theta=0,1$ we get
\begin{equation}
\rho=-p(-1)^{q}\frac{1}{2\pi a^{3}}\zeta(3),
\end{equation}
which is the Casimir energy with periodic boundary condition. For a escalar field we get the standard result. For the fermion field we have $(p,q)=(2,1)$ and we have that the energy is positive\begin{equation}
\rho=\frac{1}{\pi a^{3}}\zeta(3),
\end{equation}
which coincides with the result found in Ref. \cite{Suzuki}. For anti-periodic boundary condition $\theta=1/2$ we again find the results of Refs. \cite{Suzuki,Feng:2013zza}. From now on we will consider the general case (\ref{CasimirScalar3D}). We can see from the above result that we just get the $\lambda$ term and therefore it is not enough to generate our transversable wormhole. In the next section we will consider some possibilities.

\subsection{Massive Fields}
In order to get the constant density we first try to introduce mass to our fields. The only difference with the massless  case is that now we have
$$
\rho=p(-1)^{q}\frac{\pi^{\frac{d+1}{2}}}{a^{d+1}}\frac{\Gamma(\frac{s+1-d}{2})}{\Gamma(\frac{s}{2})}\sum_{-\infty}^{\infty}\left[(n+\theta)^{2}+(\frac{ma}{2\pi})^{2}\right]^{\frac{d-1-s}{2}}.
$$
Again, we will follow a different path than that used in Ref. \cite{Suzuki}. If we use  the Epstein zeta defined by Eq. (\ref{Epstein}),we get that our energy density becomes
$$
\rho=-p(-1)^{q}\frac{\pi^{\frac{d}{2}}}{a^{d+1}}\frac{\Gamma(-\frac{d}{2})}{2}E(c,q=(\frac{ma}{2\pi})^2,\frac{s+1-d}{2}).
$$
Now, performing the same procedures as before and using
$$
A=2,l=-d/2,c=\theta,q=(\frac{ma}{2\pi})^{2},
$$
we get
$$
\rho=-p(-1)^{q}\frac{\Gamma(-\frac{d+1}{2})}{2^{d+2}\pi^{\frac{d+1}{2}}}m{}^{d+1}-p(-1)^{q}\frac{m^{\frac{d+1}{2}}}{a^{\frac{d+1}{2}}}\frac{1}{2^{\frac{d-3}{2}}\pi^{\frac{d+1}{2}}}\sum_{n\neq0}\cos(2\pi n\theta)n^{-\frac{d+1}{2}}K_{\frac{d+1}{2}}\left(nma)\right).
$$
We should point out that for small arguments we have
$$
K_{\frac{d+1}{2}}\approx\frac{\Gamma(\frac{d+1}{2})}{2}\left(\frac{2}{z}\right)^{\frac{d+1}{2}},
$$
and in this situation, the above expression reduces to our massless case given by Eq.(\ref{casimirgeneral}). For $d=2$ we get
\begin{equation}\label{casimirmass}
\rho=-p(-1)^{q}\frac{1}{12\pi}m{}^{3}-p(-1)^{q}\frac{m^{\frac{3}{2}}}{a^{\frac{3}{2}}}\frac{\sqrt{2}}{\pi^{\frac{3}{2}}}\sum_{n=1}^{\infty}\cos(2\pi n\theta)n^{-\frac{3}{2}}K_{\frac{3}{2}}(nma).
\end{equation}
At this point we present some comments about the results expressed above. At first sight we could think that the first term would provide us with the constant density we need to the Casimir wormhole. However, the sum depends on the mass and should be expanded up to order $m^{3}$. Note that in the $(3+1)$ dimensional case, we can expand the Bessel function up to order $m^{2}$ and the sum will be convergent. This gives the usual small mass limit. However, in the $(2+1)$ dimensional case, if we expand the Bessel function, the sum converges only up
to $m^{0}$. Therefore, in $(2+1)-$ D, the above expression is not suited to consider mass corrections. In order to get this we must expand our original expression to get
\begin{eqnarray}
\rho&=&p(-1)^{q}\frac{\pi^{\frac{d+1}{2}}}{a^{d+1}}\frac{\Gamma(\frac{s+1-d}{2})}{\Gamma(\frac{s}{2})}\sum_{-\infty}^{\infty}\left[(n+\theta)\right]^{d-1-s}+p(-1)^{q}\frac{\pi^{\frac{d+1}{2}}}{a^{d+1}}\frac{\Gamma(\frac{s+1-d}{2})}{\Gamma(\frac{s}{2})}(\frac{ma}{2\pi})^{2}\frac{d-1-s}{2}\nonumber\\
&\times&\sum_{-\infty}^{\infty}\left[(n+\theta)^{2}\right]^{\frac{d-3-s}{2}}.
\end{eqnarray}
The first term of the previous equation for $\rho$ is the Casimir energy density for the massless case, as should be expected. The second term can be expressed using the  Epstein function, which results in the following
$$
\rho=E_{cas}^{m=0}+p(-1)^{q}\frac{\pi^{\frac{d+1}{2}}}{a^{d+1}}\frac{\Gamma(\frac{s+1-d}{2})}{\Gamma(\frac{s}{2})}(\frac{ma}{2\pi})^{2}\frac{d-1-s}{2}E(c=\theta,q=0,\frac{s-3-d}{2}).
$$
Performing the same procedure as before we find that the second term is null. Therefore, our Casimir energy is give by
$$
\rho=E_{cas}^{m=0}+{\cal O}(m^{4}).
$$
We should point that higher order corrections would give us terms $a^{k}$ with $k>1$ and this do not solve our problem. Therefore, only the addition of mass can not solve our problem. In the next section we will show that the application of a perpendicular magnetic field, in a graphene sheet, can provides the source to solve our problem.

\section{Casimir wormhole in a graphene sheet under a uniform magnetic field}

In this section we consider the application of the previously discussed features concerning 3-D Casimir wormholes to a graphene sheet. According to \cite{Suzuki}, the Casimir energy density of a massless fermionic field on the graphene at zero temperature is given by
\begin{equation}\label{DensEn}
\rho_C(r)=-\frac{3\zeta(3)\hbar v_F}{16 \pi r^3},
\end{equation}
considering anti-periodic boundary conditions for the field. Otherwise, the Casimir energy density obtained from periodic boundary conditions, which is positive, does not generate wormholes, since the flare out condition is not satisfied. Here we make $c\to v_F$, which is the Fermi velocity, associated to the carriers in graphene ($v_F\approx 10^3$ km$/$s) at 0 K.

The Casimir radial tension is given by
\begin{equation}\label{RadTen}
\tau_C(r)=-\frac{3\zeta(3)\hbar v_F}{8\pi r^3},
\end{equation}
so that the EoS is $\tau_C=2\rho_C$.
As the graphene sheet is immersed in a $(3+1)$ dimensional space, we get the interesting possibility of applying a magnetic field perpendicular to it. According to \cite{Correa}, this adds the term $-(+) e B m^*v_F^2/2\pi\hbar $ to $\rho_C(\tau_C)$ in Eqs. (\ref{DensEn}) and (\ref{RadTen}), with $e$ being the electron charge and $m^*$ its effective mass. Therefore, the first order corrections  to the Casimir energy density (radial tension) in presence of a uniform perpendicular magnetic field, is given by
$$
\lambda_0=- e B m^*v_F^2/2\pi\hbar
$$
This is exactly our solution with $\lambda_1=0$. In what follows, we will determine both the shape and redshift functions concerning the graphene wormhole, from the corrected energy density and tension. We find, therefore
\begin{eqnarray}
M(r)&=&-\frac{6\zeta(3)\ell}{b_0}+\frac{6\zeta(3) \ell}{r}-\frac{4 e \overline{B} v_F m^{*} \ell r^2}{\hbar^2}, \nonumber\\
2\Phi(r)&=&-\frac{r}{b_0}-\frac{[b_0+6\zeta(3)\ell]}{b_0^3}r^2+\log{\left(\frac{b_0}{r}\right)^2}.
\end{eqnarray}
where the integration constant was fixed in order to leave the logarithm argument without dimension and $\ell=\hbar G_{eff}/v_F^3$ is a characteristic length associated to the material, with $G_{eff}$ being the effective gravitational constant on the sheet. Thus, $\Lambda=-4 e \overline{B} v_F m^{*} \ell/\hbar^2$ in order to the wormhole to be asymptotically flat (in fact, conical).  We also must adjust the applied  magnetic field exactly to
\begin{equation}\label{FineTun}
 \overline{B} =\frac{3\zeta(3)\hbar^2}{2 e v_Fm^{*}b_0^3},
\end{equation}
in order that the system does not present an event horizon. With all this, the metric of the Casimir wormhole in the graphene sheet is finally given by

\begin{equation}\label{graphene metric}
ds^2=-\exp{\left[-\frac{r}{b_0}-\frac{[b_0+6\zeta(3)\ell]}{b_0^3}r^2\right]\left(\frac{b_0}{r}\right)^2}dt^2+\frac{b_0dr^2}{3\ell\zeta(3)(1-\frac{b_0}{r})} + r^2 d\phi^2.
\end{equation}

We depict in Fig. 1 the graphene Casimir wormhole, revealing the conical shape in the asymptotic limit.

\begin{figure}[H]
\centering
\includegraphics[width=0.62\textwidth]{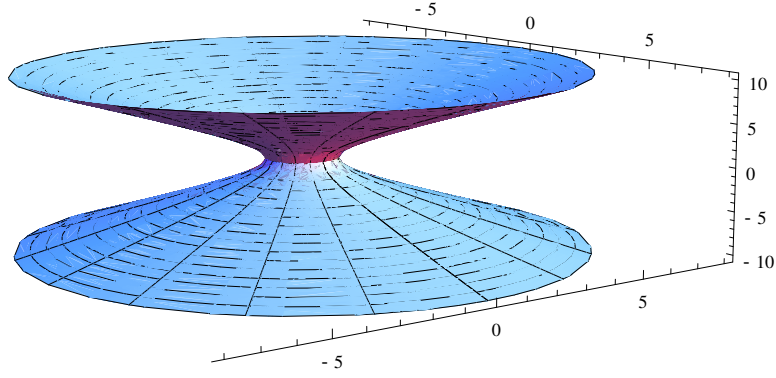}
\caption{3-D plot of a Casimir wormhole on a graphene sheet. Distances given in nm, with $r_0=1$ nm and $\ell=0.246$ nm.}
\end{figure}

Now let is examine the transport of the carriers through the wormhole, calculating the effective crossing time to go from a region at $r=-q$ to another at $r=q$ ($q\geq b_0$), given by the expression
\begin{equation}
\Delta \tau=\int^{q}_{-q} \sqrt{g_{tt}(r)}\frac{dt}{dr}dr,
\end{equation}
with $g_{tt}(r)$ given in Eq. (\ref{graphene metric}). Here, $dt/dr=(v_F)^{-1}$, and as this integral cannot be analytically solved, we depict in Fig. 2 the difference between the times, in picoseconds, which the carrier spends to run a distance $2 q$, $\Delta t$ (without the wormhole, therefore) and the one that it spends to travel the equivalent distance through the Casimir wormhole, $\Delta \tau$, both with the Fermi velocity. The parameter $\ell=2.46$ {\AA} is the lattice constant of the graphene. The graph suggests that the presence of the wormhole in the sheet represents a vantage with respect to the efficiency of the electronic transport throughout the material, better the smaller the size of the throat.
\begin{figure}[H]
\centering
\includegraphics[width=0.6\textwidth]{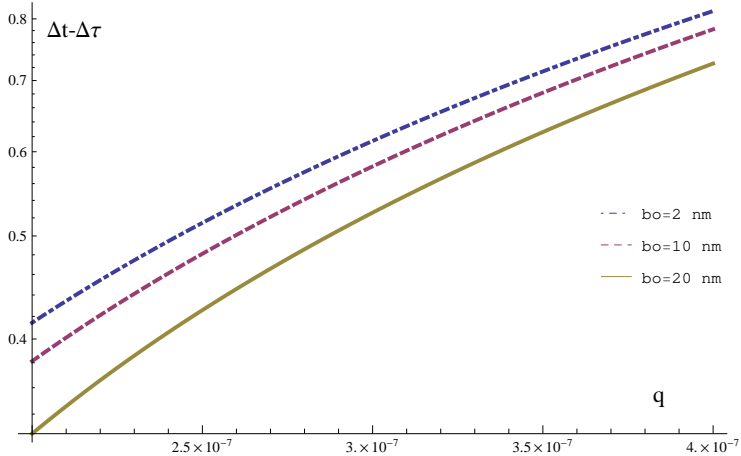}
\caption{Difference between the route times of the charge carrier by equivalent distances, in picoseconds, on a graphene sheet. The first time corresponds to a route traveled on a usual flat sheet and the second one to the radial path run through a Casimir wormhole that joins two of its regions, as a function of $q$, in meters, for the throat radii indicated in the legend and $\ell=2.46$ {\AA}.}
\end{figure}

\section{Conclusion}

In this paper we have studied 3-D traversable wormholes and shown that they cannot be sourced by only the Casimir energy density and radial tension. Recently, it has been demonstrated \cite{Remo} that in 4-D case this is possible by the direct computation of the redshift and shape functions based on a Morris-Thorne wormhole solution. However, we have presented arguments showing that in 3-D the same is not true since the arising of an event horizon is inevitable. The general analysis was made for massive and massless fermions, as well as for scalar fields, with quasi-periodic boundary conditions. We found that a possibility to circumvent the pointed out trouble is to introduce a cosmological constant, which works as an intrinsic tension on the surface, then immersing it in a 4-D (flat) manifold and applying an external tension on the surface.

We have extended the model for a graphene sheet, and obtained an asymptotically conical wormhole by specifically considering anti-periodic boundary conditions for the fermion coupled to the external magnetic field, which is source of the tension. Thus, the flare out conditions are satisfied, and adjusting the parameters we avoided the event horizon. In addition, we have investigated the electronic transport through the Casimir wormhole and shown that it is faster for a smaller wormhole throat in comparison with what happens on a flat sheet (without the wormhole), at least for some values of the effective distance travelled by the carriers. Though the difference be of only tenths of a picosecond, a charge that oscillates much times throughout the wormhole could have its comparative frequency sensibly augmented, which obviously represents a technologically attractive feature.

\section*{Acknowledgements}

The authors would like to thank Conselho Nacional de Desenvolvimento Cient\'{i}fico e Tecnol\'{o}gico (CNPq) and Funda\c{c}\~{a}o Cearense de Apoio ao Desenvolvimento Cient\'{\i}fico e Tecnol\'{o}gico (FUNCAP), under grant PRONEM PNE-0112-00085.01.00/16, for the partial financial support.

\bibliographystyle{unsrt}


\begin{thebibliography}{1}


\bibitem{EinsteineRose} Einstein, A.; and Rosen, N. The particle problem in the general theory of relativity. Phys. Rev. \textbf{48}, 73 (1935).
\bibitem{Wheeler} Misner, C. W.; and Wheeler, J. A. Classical physics as geometry: gravitation, electromagnetism, unquantized charge, and mass as properties of curved empty space. Ann. Phys. \textbf{2}, 525 (1957).
\bibitem{Ellis} Ellis, H. G. Ether flow through a drainhole - a particle model in general relativity. J. Math. Phys. \textbf{14}, 104 (1973).
\bibitem{Visser} Visser, M. Lorentzian Wormholes: From Einstein to Hawking. American Institute of Physics, New York, (1996).
\bibitem{Lemos} Dias, G. A. S. and Lemos, J. P. S., Thin-shell wormholes in d-dimensional general relativity: Solutions, properties, and stability, Phys. Rev. {\bf D 82}, 084023 (2010).
\bibitem{wormholecasimir1} Garattini, R. Casimir wormholes. Eur. Phys. J. C \textbf{79}, 951 (2019).
\bibitem{wormholecasimir2} Jusufi, K.; Channuie, P.; and Jamil, M. Traversable wormholes supported by GUP corrected Casimir energy. Eur. Phys. J. C \textbf{80}, 127 (2020).
\bibitem{wormholecasimir3} Khabibullin, A. R.; Khusnutdinov, N. R.; and Sushkov, S. V. The Casimir effect in a wormhole spacetime. Class. Quant. Grav. \textbf{23}, 627 (2006).
\bibitem{wormholecasimir4} Butcher, L. M. Casimir energy of a long wormhole throat, Phys. Rev. D {\bf 90}, 024019 (2014).
\bibitem{Maldacena1} Maldacena, J. and Milekhin, A., Humanly traversable wormholes, Phys. Rev. {\bf D 103}, 066007 (2021).
\bibitem{Geova} Lima,  A.P.C.M., Alencar, G., Muniz, C.R., and Landim, R.R., Null second order corrections to Casimir energy in weak gravitational field, JCAP 2019, 11 (2019).
\bibitem{Sorge2} Sorge, F., Quasi-local Casimir energy and vacuum buoyancy in a weak gravitational field, Class. Quantum Grav. {\bf38} 025009 (2020).
\bibitem{Calloni} Calloni, E. {\it et al.}, The Archimedes experiment, Nucl. Inst. Meth. Phys. Res., sec. A, 824, 646 (2016).
\bibitem{Sorge} Sorge, F. Casimir effect around an Ellis wormhole. Int. J. Mod. Phys. D \textbf{29}, (2019).
\bibitem{Alana} A. C. L. Santos, C. R. Muniz, and L. T. Oliveira, Casimir effect in a Schwarzschild-like wormhole spacetime, Int. J. Mod. Phys. D, DOI: 10.1142/S0218271821500322.
\bibitem{Bezerra} Bezerra, V. B., Muniz, C. R. and Toledo, J.M., Casimir Effect in spacetimes of Rotating Wormholes, Eur. Phys. J. C, DOI: 10.1140/epjc/s10052-021-09000-3.
\bibitem{Remo} Garattini, R., Casimir Wormholes, Eur.Phys. J. {\bf C79}, 951 (2019).
\bibitem{Jusufi} Jusufi, K., Channuie, P., and Jamil, M., Traversable wormholes supported by GUP corrected Casimir energy, Eur.Phys. J. {\bf C80}, 127 (2020).
\bibitem{Sunil} Tripathy, S. K., Modelling Casimir wormholes in extended gravity, Phys.Dark Univ. {\bf 31}, 100757 (2021).
\bibitem{Castro} Castro Neto, A. H. {\it et al.}, The electronic properties of graphene, Rev. Mod. Phys., {\bf 81}, 109 (2009).
\bibitem{Gonzales} Gonzalez, J. and Herrero, J., Graphene wormholes: A condensed matter illustration of Dirac fermions in curved space, Nucl. Phys. {\bf B
825} 426 (2010).
\bibitem{Furtado} Garcia, G. Q., Porfírio, P.J., Moreira, D.C., and Furtado, C., Graphene wormhole trapped by external magnetic field, Nucl.Phys.B {\bf 950},
\bibitem{Mann} Perry, G. P. and Mann, R. B., Traversible Wormholes in (2+1) Dimensions, Gen. Rel. and Grav. {\bf 24}, 305 (1992).
\bibitem{Suzuki} Ishikawa, T., Nakayama, K., and Suzuki, K., Lattice-fermionic Casimir effect and topological insulators, KEK-TH-2286,  arXiv:2012.11398v1 (2020).
\bibitem{Mann2} Delgaty, M. S. R. and Mann, R. B., Traversable Wormholes in (2+1)and (3+1) Dimensions with a Cosmological Constant, Int.J.Mod.Phys. {\bf D4}, 231 (1995).





\bibitem{Feng:2013zza}
C.~J.~Feng, X.~Z.~Li and X.~H.~Zhai,
Mod. Phys. Lett. A \textbf{29}, 1450004 (2014)
doi:10.1142/S0217732314500047
[arXiv:1312.1790 [hep-th]].
\bibitem{Elizalde} Elizalde, E.,  Zeta functions: formulas and applications, Journal of Computational and Applied Mathematics n 118,125 (2000).

\bibitem{Correa} Correa, D. H., Vacuum energies for the relativistic Landau problem, R. Number La Plata-Th 00/8,  arXiv:hep-th/0008223v1 (2000).
\end{thebibliography}

\end{document}